\documentclass[11pt,twoside]{article}

\usepackage{asp2014}

\aspSuppressVolSlug
\resetcounters

\begin{document}

\title{Feasibility of access EGI resources through the ESCAPE developed ESFRI Science Analysis Platform}

% Note the position of the comma between the author name and the 
% affiliation number.
% Authors surnames should come after first names or initials, eg John Smith, or J. Smith.
% Author names should be separated by commas.
% The final author should be preceded by "and".
% Affiliations should not be repeated across multiple \affil commands. If several
% authors share an affiliation this should be in a single \affil which can then
% be referenced for several author names. If only one affiliation, no footnotes are needed.
% See ManuscriptInstructions.pdf and ASP's manual2010.pdf 3.1.4 for more details
%G. Taffoni, S. Bertocco, M. Parra-Royón, D. Morris, K. Kliffen, F. Tinarelli, M. Stagni, V. Galluzzi
%F. Bedosti, M. Molinaro, J. Swinbank, S. Sanchez Exposito
%\author{Sample~Author1,$^1$ Sample~Author1,$^2$ and Sample~Author3$^2$}
\author{Giuliano~Taffoni,$^1$ Sara~Bertocco,\textsuperscript{1,\dag} Dave~Morris,$^2$ Manu~Parra-Roy\'{o}n,$^3$  Klaas~Kliffen,$^4$ Marco~Molinaro,$^1$ John~Swinbank,$^4$ and Susana~Sanchez Exposito$^3$}
%\affil{$^1$Institution Name, Institution City, State/Province, Country; \email{AuthorEmail@email.edu}}
\affil{\dag Corresponding author: Sara Bertocco, sara.bertocco@inaf.it}
\affil{$^1$INAF, Istituto Nazionale di Astrofisica, Italy }
\affil{$^2$University of Edinburgh}
\affil{$^3$Instituto de Astrof\'{i}sica de Andaluc\'{i}a (IAA-CSIC)}
\affil{$^4$ASTRON, Netherlands Institute for Radio Astronomy}
% This section is for ADS Processing.  There must be one line per author. paperauthor has 9 arguments.
%\paperauthor{Sample~Author1}{Author1Email@email.edu}{ORCID_Or_Blank}{Author1 Institution}{Author1 Department}{City}{State/Province}{Postal Code}{Country}
\paperauthor{Marco~Molinaro}{marco.molinaro@inaf.it}{orcid.org/0000-0001-5028-6041}{INAF}{OATs}{Trieste}{}{34143}{Italy}
\paperauthor{Giuliano~Taffoni}{giuliano.taffoni@inaf.it}{orcid.org/0000-0002-4211-6816}{Istituto Nazionale di Astrofisica}{Osservatorio Astronomico di Trieste}{Trieste}{}{34143}{Italy}
\paperauthor{Sara~Bertocco}{sara.bertocco@inaf.it}{0000-0003-2386-623X}{INAF}{OATs}{Trieste}{}{34131}{Trieste}
\paperauthor{Klaas~Kliffen}{kliffen@astron.nl}{0000-0002-4492-6732}{ASTRON}{Innovation and Systems}{Dwingeloo}{}{7991 PD}{Netherlands}
\paperauthor{John~Swinbank}{swinbank@astron.nl}{0000-0001-9445-1846}{ASTRON}{Innovation and Systems}{Dwingeloo}{}{7991 PD}{Netherlands}
\paperauthor{Dave~Morris}{dmr@roe.ac.uk}{0000-0001-6847-2328}{Institute for Astronomy}{University of Edinburgh}{Edinburgh}{}{EH93HJ}{United Kingdom}
\paperauthor{Susana~Sanchez~Exposito}{sse@iaa.es}{0000-0002-7510-7633}{Instituto de Astrof\'{i}isica de  Andaluc\'ia (IAA-CSIC)}{}{Granada}{Andaluc\'{i}ia}{18008}{Spain}
\paperauthor{Manuel~Parra-Roy\'{o}n}{mparra@iaa.es}{0000-0002-6275-8242}{IAA-CSIC}{Extragalactic Astronomy Department}{Granada}{Granada}{18008}{Spain}
% There should be one \aindex line (commented out) for each author. These are used to
% build up the author index for the Proceedings. The surname must come first, followed by
% initials. Note the use of ~ before each initial to control spacing.
% The \author entries (see above) have surname last. These \aindex entries have
% surname first.
% The Aindex.py command willl create them for you after you have constructed the \author
% The first entry should be the first author, for bold-facing the author index page-reference

%\aindex{FistAuthor1,~S.~A.}
%\aindex{Author2,~S.~B.}
%\aindex{Author3,~S.}

\begin{abstract}
The EU ESCAPE project is developing ESAP, ESFRI\footnote{ESFRI Research Infrastructures are facilities, resources or services of a unique nature, identified by European research communities to conduct and to support top-level research activities in their domains. The ESFRI Projects are Research Infrastructures in their preparation phase which have been selected for the excellence of their scientific case and for their maturity, according to a sound expectation that the Project will enter the Implementation Phase in a fixed period of time.} Scientific Analysis Platform, as an API gateway that enables the seamless integration of independent services accessing distributed data and computing resources. In ESCAPE we are exploring the possibility of exploiting EGI's OpenStack cloud computing services through ESAP. 
In our contribution we briefly describe ESCAPE and ESAP, the the use cases, the work done to automate a virtual machine creation in EGI's OpenStack cloud computing, drawbacks and possible solutions.
\end{abstract}

\section{Introduction}

In this work we consider two use cases: the SKA Data Challenges used to prepare the Astronomy and Astrophysics community to work with the data to be generated by the Square Kilometer Array (SKA), and the use of containerized Virtual Observatory services.  In our contribution, we describe the technical steps performed: we established a collaboration with the CESGA cloud site to get access to on the necessary development and test resources and we automated the creation of a Virtual Machine through the EGI FedCloud client. We automated the installation on a cloud virtual machine instance of the suitable software. We present a first standalone prototype and design the interface to provide ESAP’s users with EGI resources access.

% These lines show examples of subject index entries. At this stage these have to commented
% out, and need to be on separate lines. Eventually, they will be automatically uncommented
% and used to generate entries in the Subject Index at the end of the Proceedings volume.
% Don't leave these in! - replace them with ones relevant to your paper.
%\ssindex{FOOBAR!conference!ADASS 2019}
%\ssindex{FOOBAR!organisations!ASP}

% These lines show examples of ASCL index entries. At this stage these have to commented
% out, and need to be on separate lines. Eventually, they will be automatically uncommented
% and used to generate entries in the ASCL Index at the end of the Proceedings volume.
% The ascl.py command will scan your paper on possible code names.
% Don't leave these in! - replace them with ones relevant to your paper.
%\ooindex{FOOBAR, ascl:1101.010}

\section{The ESCAPE Project}
ESCAPE, the European Science Cluster of Astronomy \& Particle physics ESFRI research infrastructures, is an EU funded project running from 2019 through early 2023. The project aims to address the Open Science challenges shared by a range of ESFRI facilities (SKA, CTA, KM3Net, EST, ELT, HL-LHC, FAIR) and pan-European research infrastructures (CERN, ESO, JIVE) in astronomy and particle physics.
Our interest is focused on two of the main actions of the project:
\begin{itemize}
\checklistitemize
\item connect ESFRI projects to EOSC ensuring integration of data and tools 
\item establish interoperability within EOSC as an integrated multi-messenger facility for fundamental science.
\end{itemize}

\section{The ESAP platform}
ESAP, the ESFRI Science Analysis Platform, is a flexible science platform for the analysis of open access data available through the EOSC environment. Due to its modular and extensible architecture and design it can be used as a science platform toolkit for building “science platforms” which can be optimized to particular applications. It is can be deployed at a variety of scales: as a “Centralized ESAP”, providing flexible and convenient access to a wide spectrum of ESCAPE services or as a “Project ESAP”, providing a way for individual infrastructures, projects, etc to quickly integrate diverse capabilities into a unified service offering. The current prototype allows EOSC researchers
\begin{itemize}
\checklistitemize
\item to identify and stage existing data collections for analysis;
\item to collect selected data in a "shopping basket" with the same mechanism implemented in the online stores, like Amazon, where you can save in your shopping cart the products you are interested in;
\item to select among software tools and packages developed by the ESFRIs or bring their own custom workflows to the platform;
\item to exploit the underlying computing infrastructure to execute those workflows.
\end{itemize}
Researchers can select datasets from different archives.
A set of workflows Available in the ESCAPE OSSR (Open-source Scientific Software and Service Repository) is provided, the researchers can select the workflow of interest and then choose to deploy it in a suitable infrastructure.

\section{The use cases}
The workflows provided in ESAP are relative to use cases of interest for the ESFRI and big research infrastructures users. In our work we considered the need to provide users both with containerized software and with an environment allowing to run more containers, possibly also sequentially in an interactive environment. We considered specifically two use cases:\begin{itemize}
\checklistitemize
\item To provide access to containerized IVOA-enabled applications\footnote{\url{https://github.com/zarquan/Oligia}} 
\item To provide an environment to run the workflow to process HI data cubes produced by radio interferometers, in particular large data cubes produced by future instruments like the SKA\footnote{\url{https://hi-friends-sdc2.readthedocs.io/en/latest/}}\footnote{\url{https://github.com/HI-FRIENDS-SDC2/hi-friends}}.  
\end{itemize}

\section{The EGI computing infratructure}
EGI is a federation of computing and storage resource providers delivering open solutions for advanced
computing and data analytics to support research and development. The EGI Core is the EGI
Federated Cloud Platform (FedCloud), a federation and management platform that pool together various resources. It provides various compute services. The applicable to our use cases are {\bf Cloud Compute} which provides virtual machine-based computing with associated storage, providing users with the ability to deploy and scale virtual machines on-demand, and
{\bf Container Compute}, which supports running container-based applications with either Docker or Kubernetes on top of Cloud Compute. In this work, we exploited the cloud compute service and we plan to evaluate the container compute service in the future. 

\section{Automation of Virtual Machine creation in OpenStack EGI FedCloud feasibility}
To create a VM in OpenStack EGI FedCloud, following the provided documentation\footnote{\url{https://docs.egi.eu/users/getting-started}}, we registered to the piloting purpose Virtual Organisation vo.access.egi.eu; we asked CESGA site support to be able to allocate needed resources (particularly a public IP address). Then, using the EGI FedCloud Python client \footnote{\url{https://github.com/tdviet/fedcloudclient}}
we wrote a simple Djando web application, easily integrated in ESAP, to automate the virtual machine creation process in the EGI FedCloud.
\articlefigure{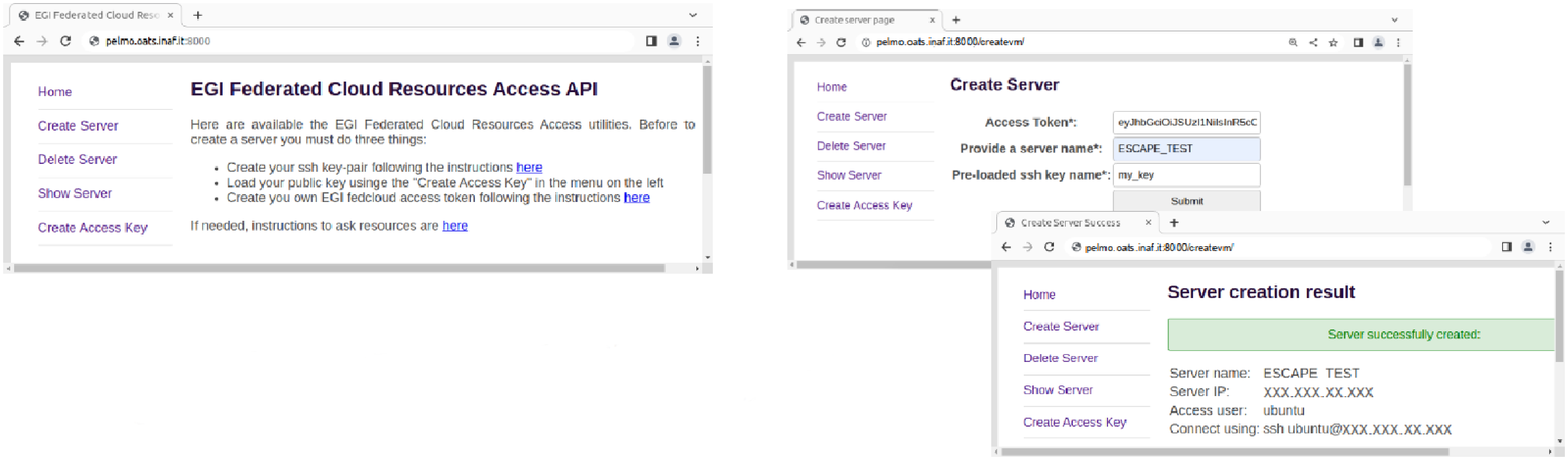}{fig2}{GUI EGI server creation}

\section{Lessons learned}
We have identified some show stoppers in the virtual machine automation process:
\begin{itemize}
\checklistitemize
\item \textbf{Public network}. Each site has a slightly different network configuration, and there is no a standard way to tell which one to use and how to setup the router for it. There is not a convention for network naming, so human intervention --- guessing --- is needed to establish which is a public network.
\item \textbf{Supported Virtual Organizations}. There is not a command in the FedCloud tool that lists the sites that support a certain virtual organisation.
\item \textbf{Flavors and images}. Again, human intervention is required to establish the resources associated to the flavors or the operating system in an image. A naming convention could be useful. 
\end{itemize}
The authentication and authorization mechanism, as is, is not smooth because ESAP A\&A is managed through InDiGo IAM and EGI A\&A is managed through EGI Check-in, morover the authentication in the virtual machine has to be done through ssh keys.
\section{Thoughts on possible solutions}
A workaround to the lack of information about supported Virtual Organizations and availabe VMs images and flavours could be the implementation of a service collecting and periodically updating the needed information through OpenStack/FedCloud client commands like 

\begin{description}
  \item[\texttt{openstack image list}]to list available flavors
  \item[\texttt{openstack image show <image>}] to display image details
  \item[\texttt{openstack flavor list}] to list available flavors
  \item[\texttt{openstack flavor show <flavor>}] to display flavor details
\end{description}

Authentication and authorization (A\&A) at infrastructure level could be made
smooth by exploiting the fact that InDiGo IAM and EGI Check-in are based on the same underlying technologies (OpenID Connect, with JSON Web Tokens for authentication and group
membership for authorization) and making a mapping between entities.
At low level, A\&A could be made smooth exploiting a software like
''motley cue'' mapping OIDC identities to local identities.\footnote{https://motley-cue.readthedocs.io/en/latest/} 

\acknowledgements The project leading to this application has received funding from the European Union's Horizon 2020 research and innovation programme under grant agreement No 824064".\\
Thanks to CESGA for providing resources used in our tests.

%\bibliography{example}  % For BibTex

% if we have space left, we might add a conference photograph here. Leave commented for now.
% \bookpartphoto[width=1.0\textwidth]{foobar.eps}{FooBar Photo (Photo: Any Photographer)}

\end{document}